\documentclass[prd,showpacs,twocolumn,preprintnumbers,amsmath,amssymb,superscriptaddress,floatfix,nofootinbib]{revtex4-1}%

\usepackage{graphicx}
\usepackage{bm}
\usepackage{amsmath}
\usepackage{amsfonts}
\usepackage{amssymb}
\usepackage{color}
\usepackage{multirow}
\usepackage{subfigure}
\usepackage[colorlinks, citecolor=blue,anchorcolor=red,menucolor=red, linkcolor=red,filecolor=red,runcolor=red,urlcolor=blue,frenchlinks=red]{hyperref}

\begin{document}
	\title{The $\Lambda_c^+\to\eta\pi^+\Lambda$ reaction and the $\Lambda a_0^+(980)$ and $\pi^+\Lambda(1670)$ contributions}

	\author{Man-Yu Duan}\email{duanmy@seu.edu.cn}
	\affiliation{Department of physics, Guangxi Normal University, Guilin 541004, China}
	\affiliation{Guangxi Key Laboratory of Nuclear Physics and Nuclear Technology, Guangxi Normal University, Guilin 541004, China}
	\affiliation{School of Physics, Zhengzhou University, Zhengzhou 450001, China}
	\affiliation{School of Physics, Southeast University, Nanjing 210094, China}
	\vspace{0.5cm}
	
	\author{Wen-Tao Lyu}\email{wentaolyu@gs.zzu.edu.cn}
	\affiliation{School of Physics, Zhengzhou University, Zhengzhou 450001, China}
	\affiliation{Department of physics, Guangxi Normal University, Guilin 541004, China}\vspace{0.5cm}
	
	\author{Chu-Wen Xiao}\email{xiaochw@gxnu.edu.cn}
	\affiliation{Department of physics, Guangxi Normal University, Guilin 541004, China}
	\affiliation{Guangxi Key Laboratory of Nuclear Physics and Nuclear Technology, Guangxi Normal University, Guilin 541004, China}
	\affiliation{School of Physics, Central South University, Changsha 410083, China}\vspace{0.5cm}
	
	\author{En Wang}\email{wangen@zzu.edu.cn}
	\affiliation{School of Physics, Zhengzhou University, Zhengzhou 450001, China}\vspace{0.5cm}
	
	\author{Ju-Jun Xie}\email{xiejujun@impcas.ac.cn}
    \affiliation{Southern Center for Nuclear-Science Theory (SCNT), Institute of Modern Physics, Chinese Academy of Sciences, Huizhou 516000, China}
    \affiliation{Heavy Ion Science and Technology Key Laboratory, Institute of Modern Physics, Chinese Academy of Sciences, Lanzhou 730000, China} 
	\affiliation{School of Nuclear Sciences and Technology, University of Chinese Academy of Sciences, Beijing 101408, China}

	\author{Dian-Yong Chen}\email{chendy@seu.edu.cn}
	\affiliation{School of Physics, Southeast University, Nanjing 210094, China}
	\affiliation{Lanzhou Center for Theoretical Physics, Lanzhou University, Lanzhou 730000,China}
	
	\author{Eulogio Oset}\email{oset@ific.uv.es}
	\affiliation{Department of physics, Guangxi Normal University, Guilin 541004, China}
	\affiliation{Departamento de Física Teórica and IFIC, Centro Mixto Universidad de Valencia-CSIC Institutos de Investigación de Paterna, 46071 Valencia, Spain}
	\affiliation{School of Physics, Zhengzhou University, Zhengzhou 450001, China}\vspace{0.5cm}
\begin{abstract}
We study from the theoretical point of view the $\Lambda_c^+\to \pi^+ \eta \Lambda$ reaction, recently measured by the Belle and BESIII Collaborations, where clear signals are observed for $a_0(980)$, $\Lambda(1670)$, and $\Sigma(1385)$ excitation. By considering the $a_0(980)$ and $\Lambda(1670)$ as dynamically generated resonances from the meson meson and meson baryon interaction, respectively, we are able to determine their relative production strength in the reaction, which is also tied to the strength of the $\pi^+ \eta \Lambda$ tree level contribution. We observe that this latter strength is very big and there are large destructive interferences between the tree level and the rescattering terms where the  $a_0(980)$ and $\Lambda(1670)$ are generated. The $\Sigma(1385)$ contribution is included by means of a free parameter, the only one of the theory, up to a global normalization, when one considers only external emission, and we observe that the spin flip part of this term, usually ignored in theoretical and experimental works, plays an important role determining the shape of the mass distributions. Internal emission is also considered and it is found to play a minor role.
\end{abstract}
	
	\date{\today}
	
	\maketitle
	
\section{Introduction}\label{sec1}

The $\Lambda_c^+\to\Lambda\pi^+\eta$ reaction has been measured with precision by the Belle Collaboration~\cite{Belle:2020xku} and two clear structures are stressed: One peak in the $\eta\Lambda$ mass distribution, which is associated with the $\Lambda(1670)(1/2^-)$ resonance, and another peak in the $\pi^+\Lambda$ mass distribution associated with the $\Sigma(1385)(3/2^+)$ resonance. Interestingly, the $\Lambda a_0^+(980)$ mode, which one expects from the $\pi^+\eta$ coming from the $a_0^+(980)$, is seen but with very small intensity compared to the $\Sigma(1385)$ (see diagonal in the Dalitz plot of Fig.5 of Ref.~\cite{Belle:2020xku}). The reaction has been posteriorly measured by the BESIII Collaboration~\cite{BESIII:2024mbf} and a wider range of masses in the three invariant mass distributions, $\pi^+\Lambda$, $\pi^+\eta$, $\eta\Lambda$ is presented, together with an analysis of the data. The Dalitz plot is not shown, however, the signal for the $\Lambda a_0^+(980)$ is as faint as in the Belle measurements~\cite{rui}. In spite of the small signal seen in the Dalitz plot, the analysis of Ref.~\cite{BESIII:2024mbf} reports a branching fraction of 54\% for the $\Lambda a_0^+(980)$ decay mode. This surprising conclusion deserves a detailed attention which is the purpose of the present work, among others.

The issue has attracted much attention. In Ref.~\cite{Yu:2020vlt} a significant contribution to $\Lambda_c^+\to\Lambda a_0^+(980)$ is claimed from a triangle mechanism where $\Lambda_c^+\to \Sigma^{*+}\eta$, followed by $\Sigma^{*+}\to\Lambda\pi^+$ and final fusion of $\eta\pi^+$ to give the $a_0(980)$. However, it is easy to see that the triangle is very far away from developing a triangle singularity (see Eq.~(18) of Ref.~\cite{Bayar:2016ftu}), which indicates that this production mode is relatively inefficient. In Ref.~\cite{Wang:2022nac}, a different perspective is taken, and starting from the basic decay at the quark level with external emission, $cud\to W(u\bar{d})sud$, two modes of hadronization into mesons were considered, $u\bar{d}$ hadronization into two mesons and $su\to\sum_is\bar{q}_iq_iu$ hadronization into a meson and $\Lambda$. The latter mode produced indeed the $\Lambda\eta\pi^+$ final state. However, the first mode, where the $u\bar{d}$ is hadronized requires a close scrutiny. Indeed, the hadronization of $u\bar{d}$ gives rise to (see Ref.~\cite{Duan:2024okk})
\begin{eqnarray}
	u\bar{d} \to\left(\frac{\pi^0}{\sqrt{2}}+\frac{\eta}{\sqrt{3}}\right)\pi^++\pi^+\left(-\frac{\pi^0}{\sqrt{2}}+\frac{\eta}{\sqrt{3}}\right)+K^+\bar{K}^0,
\end{eqnarray}
and in Ref.~\cite{Wang:2022nac} the $\eta\pi^+$ and $\pi^+\eta$ components are added and the final state interaction of this channel and the $K^+\bar{K}^0$ transition to $a_0^+(980)$ are considered.

The subtle point is that, as discussed in detail in Ref.~\cite{Duan:2024okk}, the $\eta\pi^+$ and $\pi^+\eta$ terms do not add but cancel, due to the $[P,\partial_\mu P]W^\mu$ structure of the $WPP$ vertex~\cite{Gasser:1983yg,Scherer:2002tk,Ren:2015bsa}, where $P$ is the matrix of the pseudoscalar mesons. Similarly, since $\mu=0$ gives the dominant contribution, the $K^+\partial_0\bar{K}^0-\bar{K}^0\partial_0 K^+$ in the $K^+\bar{K}^0$ term also does not contribute~\cite{Duan:2024okk}. 

There is another point that we shall consider later on. While in Ref.~\cite{Xie:2017xwx} the full relativistic propagator of the $\Sigma(1385)$ is taken into account, in Ref.~\cite{Wang:2022nac} only the spin independent part of the $\Sigma^*$ propagator is considered. We shall see that the spin flip part of this amplitude plays an important role. The same approach is followed in Ref.~\cite{Lyu:2024qgc}, trying to extract a signal for the $\Sigma(1380)(1/2^-)$ production, suggested in Refs.~\cite{Wu:2009tu,Wu:2009nw}, from the analysis of the $K^-p\to\Lambda\pi^+\pi^-$ reaction (see also recent review on the low lying $\Sigma^*(1/2^-)$ state in Ref.~\cite{Wang:2024jyk}).

In Ref.~\cite{Zhang:2024jby}, where emphasis is made on the production of the $\Lambda(1670)$ in the $\Lambda_c^+\to\pi^+ K^-p$ reaction, based on the large signal reported by the BESIII Collaboration for the $\Lambda_c^+\to\Lambda a_0^+(980)$ decay channel~\cite{BESIII:2024mbf}, a triangle mechanism $\Lambda_c^+\to\Lambda a_0^+$ followed by $a_0^+\to\pi^+\eta$ and further fusion of $\eta\Lambda$ to give the $\Lambda(1670)$ state is proposed. The strength is tied to the branching fraction of $\Lambda_c^+\to\Lambda a_0^+$ of Ref.~\cite{BESIII:2024mbf}, which we try to reanalyze here.

As we can see, the interpretations of many recent reactions are tied to the issue of the $\Lambda_c^+\to\Lambda a_0^+(980)$ decay mode, and a detailed analysis of the particular $\Lambda_c^+\to\Lambda a_0^+(980)$ decay mode is called for, which we carry out in the present work.

\section{Formalism}\label{sec2}

\begin{figure}
	\centering
	\includegraphics[scale=0.65]{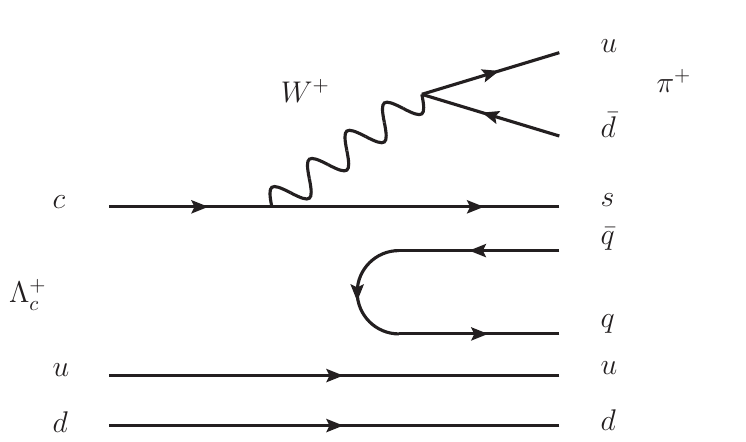}
	\caption{External emission mechanism, $cud\to\pi^+sud$, together with the hadronization of the strange quark and the $ud$ quark pair.}\label{3-body-external}
\end{figure}

As discussed in the Introduction and detailed in Ref.~\cite{Duan:2024okk}, where the $\Lambda_c^+\to pK^-\pi^+$ reaction near the $\eta\Lambda$ threshold was studied, we start from the external emission mechanism of Fig.~\ref{3-body-external} with the hadronization of the strange quark and the $ud$ quark pair.

Analytically, we have 
\begin{eqnarray}\label{Lambdac-hadronization}
	&\Lambda_c^+&=\frac{1}{\sqrt{2}}c(ud-du)\chi_{MA}\rightarrow\pi^+\frac{1}{\sqrt{2}}s(ud-du)\chi_{MA} \nonumber\\
	&=&\pi^+\sum_i\frac{1}{\sqrt{2}}s\bar{q}_iq_i(ud-du)\chi_{MA}\nonumber\\
	&=&\frac{1}{\sqrt{2}}\pi^+\sum_iP_{3i}q_i(ud-du)\chi_{MA} \nonumber\\
	&=&\frac{1}{\sqrt{2}}\pi^+\left\{K^-(uud-udu)+\bar{K}^0(dud-ddu) \right.\nonumber\\
	&&\left.-\frac{\eta}{\sqrt{3}}(sud-sdu)\right\},
\end{eqnarray}
where $\chi_{MA}$ is the mixed antisymmetric spin of the $(ud-du)$ pair, and $P$ is the SU(3) matrix of the octet of pseudoscalar mesons, which accounts for $q_i\bar{q}_j$ written in terms of the mesons,
\begin{eqnarray}
	P =\left(\begin{matrix}   \frac{{\eta}}{\sqrt{3}}+ \frac{{\pi}^0}{\sqrt{2}} +\frac{{\eta}'}{\sqrt{6}}  & \pi^+  & K^{+}  \\
		\pi^-  &    \frac{{\eta}}{\sqrt{3}}- \frac{{\pi}^0}{\sqrt{2}} +\frac{{\eta}'}{\sqrt{6}}  &  K^{0} \\
		K^{-}  &  \bar{K}^{0}   & \sqrt{\frac{2}{3}}{\eta}'  -\frac{{\eta}}{\sqrt{3}}
		\end{matrix}
	\right).
\end{eqnarray}

In order to write the three quarks state in term of baryons, we use the results of Table III of Ref.~\cite{Miyahara:2016yyh}, where the flavor mixed antisymmetric form of the octet of baryons, $\phi_{MA}$, is shown. Recalling that for the octet of baryons the wave function is $1/\sqrt{2}(\phi_{MS}\chi_{MS}+\phi_{MA}\chi_{MA})$, with $\phi_{MS}$, $\chi_{MS}$ the mixed symmetric flavor and spin wave functions, and using the results of Ref.~\cite{Miyahara:2016yyh}, Eq.~(\ref{Lambdac-hadronization}) takes the form,

\begin{equation}\label{Eq:H}
	H=\pi^+\left\{\frac{1}{\sqrt{2}}K^-p+\frac{1}{\sqrt{2}}\bar{K}^0n+\frac{1}{3}\eta\Lambda\right\},
\end{equation}
which has to be considered as the tree level contribution in $\Lambda_c^+\to H \to BPP$, with $B$ a baryon of the octet.

We can see that the last term in Eq.~(\ref{Eq:H}) is $\frac{1}{3}\pi^+\eta\Lambda$, which directly corresponds to the desired final state. This means that we have already the $\pi^+\eta\Lambda$ final state of the tree level, a feature of relevance in the interpretation of the experimental data of Ref.~\cite{BESIII:2024mbf}. However, it is also important to note that the $\bar{K}N$ channels are not innocuous, since one can have the transition $\bar{K}N\to\eta\Lambda$ and reach the $\pi^+\eta\Lambda$ final state. Indeed, $\eta\Lambda$ is one of the terms of the coupled channels in the $K^-p$ interaction accounting for the meson baryon interaction of the octet of mesons with the octet of baryons which are considered in the chiral unitary approach~\cite{Kaiser:1995eg,Oset:1997it,Oller:2000fj,Garcia-Recio:2002yxy,Oset:2001cn,Oller:2006jw,Jido:2003cb,Cieply:2016jby,Oller:2000ma,Hyodo:2011ur,Meissner:2020khl}.

In addition to the $\bar{K}N\to\eta\Lambda$ we must also consider the $\eta\Lambda\to\eta\Lambda$ and $\eta\pi^+\to\eta\pi^+$ rescattering, the latter reaction also been considered in the chiral unitary approach for meson meson interaction in Refs.~\cite{Oller:1997ti,Kaiser:1998fi,Markushin:2000fa,Nieves:1998hp}. All the rescattering transitions disclosed above are produced in $S$-wave and we evaluate them using the chiral unitary approach. In addition, the $\pi^+\Lambda$ state can also come from $\Sigma(1385)(3/2^+)$ decay and we must take it into account. This term is introduced explicitly, while the $a_0^+(980)$ and $\Lambda(1670)$ resonances, which also show up in the experimental data, are dynamically generated in our approach and their strength is tied to the one of the $\Lambda\pi^+\eta$ tree level contribution.

\begin{figure}[htbp]
	\centering
	\includegraphics[scale=0.45]{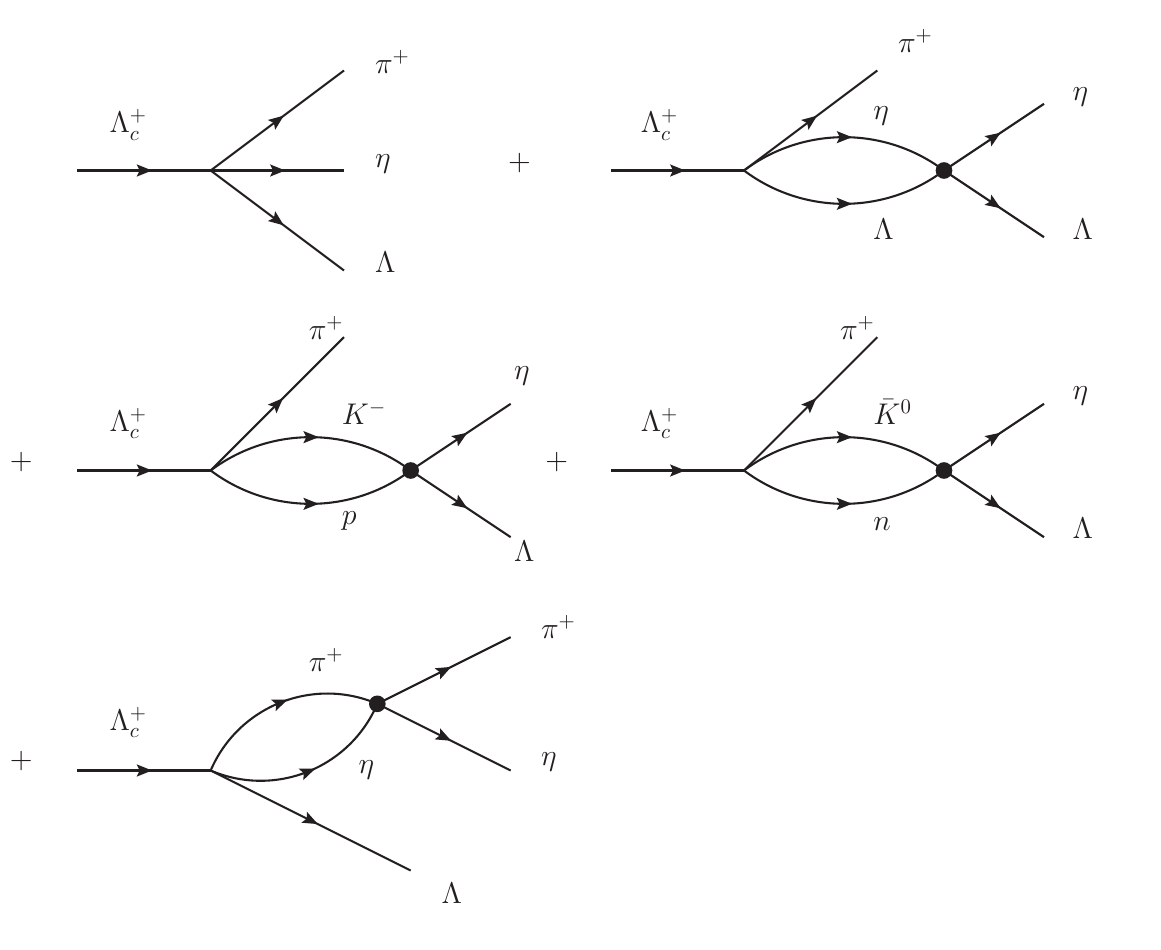}
	\caption{ Mechanisms for tree level $\Lambda_c^+\to\pi^+\eta\Lambda$ and rescattering of intermediate components.}\label{fig:S-wave}
\end{figure}

In practical terms this means that we do not include any free parameters to account for these two resonances, but the $P$-wave excited resonance $\Sigma(1385)$ is not tied to the $S$-wave resonance and it is introduced in our scheme by means of one free parameter that is fitted to the data. Thus, up to a global normalization, our results for the mass distributions depend on just one parameter, presenting a real challenge to reproduce all the mass distributions. One might think of other possible $\Sigma^*$ resonances contributing to the $\pi^+\Lambda$ invariant mass, but the experiment does not show any sizable contribution of them, which can be understood theoretically . Indeed, one generates the $\Sigma(1620)(1/2^-)$ dynamically in our approach, but the coupling to the $\pi\Lambda$ is very weak, reflected also experimentally by the 9\% branching ratio of decay to this channel. Similarly the $\Sigma(1580)(3/2^-)$ would decay into $\pi\Lambda$ in $D$-wave, also weakened.

\begin{figure}[htbp]
	\centering
	\includegraphics[scale=0.65]{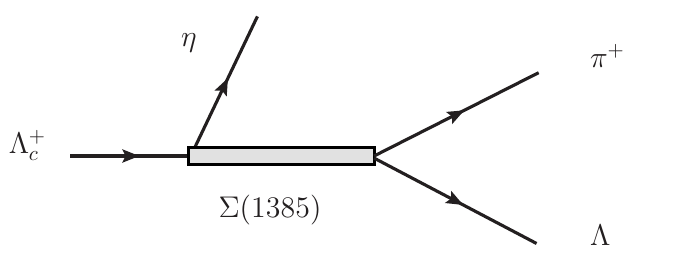}
	\caption{ Mechanism for $\Sigma(1385)$ excitation.}\label{fig:mechanism-Sigma1385}
\end{figure}

This said, diagrammatically we have the mechanisms depicted in Fig.~\ref{fig:S-wave}, which contain the tree level and $\eta\Lambda\to\eta\Lambda$, $K^-p\to\eta\Lambda$, $\bar{K}^0n\to\eta\Lambda$, and $\pi^+\eta\to\pi^+\eta$ rescattering. In addition, we have the mechanism of Fig.~\ref{fig:mechanism-Sigma1385} for $\Sigma(1385)$ excitation.

\subsection{$\Sigma(1385)(3/2^+)$ excitation}

The mechanism of $\Sigma(1385)$ excitation of Fig.~\ref{fig:mechanism-Sigma1385}, requires making a transition from a baryon state of $1/2^+$ to another one of $3/2^+$. This is analogous to the nucleon-$\Delta$ transition, which is done by means of the $S^\dagger$ transition spin operator commonly used in studies of pion nuclear physics with the $\Delta$-hole model~\cite{Ericson:1988gk,Oset:1981ih}. It is defined as 
\begin{equation}
	\left\langle\frac{3}{2}M\left|S^\dagger_\nu\right|\frac{1}{2}m\right\rangle=C\left(\frac{1}{2},\mathbf{1},\frac{3}{2}; m,\nu,M\right)\left\langle\frac{3}{2}\left|S^\dagger\right|\frac{1}{2}\right\rangle,
\end{equation}
with $\nu$ the spherical component of the rank $\mathbf{1}$ operator $\vec{S}^\dagger$, which makes explicit use of the Wigner-Eckart theorem, and the reduced matrix element is taken equal $\mathbf{1}$ as a definition. For practical calculations one makes use of 
\begin{equation}\label{Eq:S}
	\sum_M S_l|M\rangle\langle M|S^\dagger_m=\frac{2}{3}\delta_{lm}-\frac{i}{3}\epsilon_{lms}\sigma_s,
\end{equation}
with $l$, $m$, $s$ cartesian components. In Ref.~\cite{Wang:2022nac} the $\sigma$ term in Eq.(\ref{Eq:S}) is not considered, and in the analysis of Ref.~\cite{BESIII:2024mbf} we shall also see that it has not been considered either, and it plays a relevant role in the $\eta\Lambda$ mass distribution.

The amplitude for $\Sigma(1385)$ excitation, with a $P$-wave coupling in each of the $1/2\to3/2$ transition vertices, is given by

\begin{eqnarray}
	t_{\Sigma(1385)}&=&\alpha \left\langle m^\prime\left|S_i\vec{P}^*_{\pi^+ i}\right|M\right\rangle\left\langle M\left|S_j^\dagger\vec{P}_{\eta j}^*\right|m\right\rangle D  \nonumber \\
	&=&\alpha D\left\langle m^\prime\left|\frac{2}{3}\delta_{ij}-\frac{i}{3}\epsilon_{ijs}\sigma_s\right|m\right\rangle \vec{P}^*_{\pi^+ i}\vec{P}_{\eta j}^* \\
	&=&\alpha D\left\langle m^\prime\left|\frac{2}{3}\vec{P}^*_{\pi^+}\cdot\vec{P}_{\eta}^*-\frac{i}{3}\epsilon_{ijs}\sigma_s\vec{P}^*_{\pi^+ i}\vec{P}_{\eta j}^*\right|m\right\rangle,  \nonumber 
\end{eqnarray}
where 
\begin{equation}
	D=\dfrac{1}{M_{\text{inv}}(\pi^+\Lambda)-M_{\Sigma(1385)}+{i\Gamma_{\Sigma(1385)}}/{2}},
\end{equation}
where
\begin{equation}
	M_{\Sigma(1385)}=1382.8~\text{MeV},~~\Gamma_{\Sigma(1385)}=36.0~\text{MeV},
\end{equation}
with the momenta $\vec{P}^*_{\pi^+}$, $\vec{P}_{\eta}^*$ in the rest frame of the resonance, which means $\vec{P}^*_{\pi^+}+\vec{P}^*_{\Lambda}=0$. In this frame it is easy to see that
\begin{equation}
	\omega_\eta=\dfrac{M_{\Lambda_c^+}^2-M_{\text{inv}}^2(\pi^+\Lambda)-m_\eta^2}{2M_{\text{inv}}(\pi^+\Lambda)},
\end{equation}
and
\begin{eqnarray}
	M_{\text{inv}}^2(\pi^+\eta)&=&(P_{\pi^+}+P_\eta)^2\nonumber\\
	&=&m_{\pi^+}^2+m_\eta^2+2\omega_{\pi^+}\omega_\eta-2\vec{P}^*_{\pi^+}\cdot\vec{P}_{\eta}^*,\quad
\end{eqnarray}
from where
\begin{equation}\label{Eq:ppidotpeta}
	\vec{P}^*_{\pi^+}\cdot\vec{P}_{\eta}^*=\frac{1}{2}\left\{m_{\pi^+}^2+m_\eta^2+2\omega_{\pi^+}\omega_\eta-M_{\text{inv}}^2(\pi^+\eta)\right\},
\end{equation}
where
\begin{equation}
	\omega_{\pi^+}=\dfrac{M_{\text{inv}}^2(\pi^+\Lambda)+m_{\pi^+}^2-M_\Lambda^2}{2M_{\text{inv}}(\pi^+\Lambda)}.
\end{equation}

In this way we prepare the ground for expressing all the amplitudes in terms of the invariant masses, to finally calculate the mass distributions in terms of these invariant masses.

\subsection{Full decay amplitude}

We write the $\Lambda_c^+\to\pi^+\eta\Lambda$ decay amplitude as
\begin{equation}
	t=t_1+t_2,
\end{equation}
with
\begin{eqnarray}\label{Eq:t1}
	t_{1}&=&A \left\{ h_{\pi^+\eta\Lambda}+h_{\pi^+\eta\Lambda}G_{\eta\Lambda}(M_{\text{inv}}(\eta\Lambda))t_{\eta\Lambda,\eta\Lambda}(M_{\text{inv}}(\eta\Lambda))\right. \nonumber\\
	&&+h_{\pi^+\eta\Lambda}G_{\pi^+\eta}(M_{\text{inv}}(\pi^+\eta))t_{\pi^+\eta,\pi^+\eta}(M_{\text{inv}}(\pi^+\eta))  \nonumber\\
	&&+h_{\pi^+\bar{K}N}G_{K^-p}(M_{\text{inv}}(\eta\Lambda))t_{K^-p,\eta\Lambda}(M_{\text{inv}}(\eta\Lambda))   \nonumber\\
	&&+h_{\pi^+\bar{K}N}G_{\bar{K}^0n}(M_{\text{inv}}(\eta\Lambda))t_{\bar{K}^0n,\eta\Lambda}(M_{\text{inv}}(\eta\Lambda)) \nonumber\\
	&&\left.+\frac{\beta}{M_\Lambda}\frac{2}{3}\vec{P}^*_{\pi}\cdot\vec{P}_{\eta}^* D\right\},
\end{eqnarray}
where
\begin{equation}
	h_{\pi^+\eta\Lambda}=\frac{1}{3};~~~h_{\pi^+\bar{K}N}=\frac{1}{\sqrt{2}},
\end{equation} 
\begin{equation}\label{Eq:t2}
	t_2=-\frac{i}{3}\frac{A\beta}{M_\Lambda}\epsilon_{ijs}\sigma_s\vec{P}^*_{\pi i}\vec{P}_{\eta j}^* D,
\end{equation}
and $A$ is a global normalization constant and $\beta$ is a free parameter. We have chosen $\beta$ such that $A\beta/M_{\Lambda}\equiv\alpha$ for dimensional reasons. The weights $h_{\pi^+\eta\Lambda}$, $h_{\pi^+\bar{K}N}$ stem from Eq.~(\ref{Eq:H}). Taking now the sum and average over the baryon spins of $|t|^2$ and considering that the $t_{\eta\Lambda,\eta\Lambda}$, $t_{\bar{K}N,\bar{K}N}$, and $t_{\pi^+\eta,\pi^+\eta}$ amplitudes are spin independent, we obtain,
\begin{eqnarray}\label{Eq:t-total}
	\overline{\sum}\sum|t|^2&=&|t_1|^2+\frac{1}{9}\left(\frac{A\beta}{M_\Lambda}\right)^2\left|D\right|^2\nonumber\\
	&&\cdot\left\{\vec{P}^{*2}_{\pi}\cdot\vec{P}_{\eta}^{*2}-(\vec{P}^*_{\pi}\cdot\vec{P}_{\eta}^*)^2\right\}.
\end{eqnarray}

According to the notation 1 to $\pi^+$, 2 to $\eta$ and 3 to $\Lambda$ and applying the standard formula of the Review of Particle Physics (RPP)~\cite{ParticleDataGroup:2024cfk}, we have
\begin{equation}
	\dfrac{d^2\Gamma}{dM_{12}dM_{23}} = \frac{1}{(2\pi)^3}\dfrac{2M_\Lambda 2M_{\Lambda_c^+}}{32M_{\Lambda_c^+}^3}\overline{\sum}\sum|t|^2~2M_{12}~2M_{23},
\end{equation}
with the Mandl and Shaw normalization of the meson and baryon fields~\cite{Mandl:1985bg}.

We can get $d\Gamma/dM_{12}$ by integrating $d^2\Gamma/(dM_{12}dM_{23})$ over $M_{23}$ with the limits in the RPP~\cite{ParticleDataGroup:2024cfk}. Permutation of the indices allows us to evaluate all three mass distributions, using $M_{12}$, $M_{23}$ as independent variables, and the property $M_{12}^2+M_{13}^2+M_{23}^2=M_{\Lambda_c^+}^2+m_{\pi^+}^2+m_\eta^2+M_\Lambda^2$ to get $M_{13}$ from them.

\subsection{Amplitudes from the chiral unitary approach} 

The amplitudes $t_{\eta\Lambda,\eta\Lambda}$, $t_{K^-p,\eta\Lambda}$, $t_{\bar{K}^0n,\eta\Lambda}$, we take from the chiral unitary approach with the $K^-p$, $\bar{K}^0n$, $\pi^+\Sigma^-$,  $\pi^-\Sigma^+$, $\pi^0\Sigma^0$, $\pi^0\Lambda$, $\eta\Lambda$, $\eta\Sigma^0$, $K^+\Xi^-$, $K^0\Xi^0$ coupled channels of Ref.~\cite{Oset:1997it}, with the input slightly changed in Ref.~\cite{Duan:2024okk} to reproduce the precise peak of the $K^-p$ mass distribution of the $\Lambda_c^+\to\pi^+K^-p$ decay at the $\eta\Lambda$ threshold, measured with high precision by the Belle Collaboration in Ref.~\cite{Belle:2022cbs}. That work allowed us to get a combined description of the Belle data and the $K^-p\to \eta\Lambda$ data and extract from there the $\eta\Lambda$ scattering length and effective range, plus the position of the $\Lambda(1670)$ resonance with high precision. Since one of the peaks observed in the BESIII experiment corresponds to the $\Lambda(1670)$ excitation, the use of the updated input of Ref.~\cite{Duan:2024okk} is most opportune.

Concerning the $t_{\eta\pi^+,\eta\pi^+}$ amplitude we use the chiral unitary approach with the channels $K^+\bar{K}^0$ (1) and $\pi^+\eta$ (2). The transition potential is given by
\begin{eqnarray}
	V_{11}&=&-\frac{s}{4f^2},~~~~(f=93~\text{MeV}),\nonumber\\
	V_{12}&=&-\frac{1}{3\sqrt{3}f^2}(3s-2m_K^2-m_\eta^2),\nonumber\\
	V_{22}&=&-\frac{2m_\pi^2}{3f^2},
\end{eqnarray} 
obtained from Ref.~\cite{Lin:2021isc}, where the $\eta-\eta^\prime$ mixing of Ref.~\cite{Bramon:1992kr} is used\footnote{These differ only slightly from those used in Ref.~\cite{Molina:2023oeu}, where the mixing was not considered.}. And the $T$ matrix is given by 
\begin{equation}
	T=[1-VG]^{-1}V,
\end{equation}
with G the ordinary meson meson loop function
\begin{equation}
	G=\int\frac{d^3q}{(2\pi)^3}\dfrac{\omega_1(q)+\omega_2(q)}{2\omega_1(q)\omega_2(q)}\dfrac{1}{s-\left[\omega_1(q)+\omega_2(q)\right]^2+i\epsilon},
\end{equation}
with $q_{\text{max}}=600$~MeV.

A small detail on the use of the $t_{\pi^+\eta,\pi^+\eta}$ amplitude is that we kill it adiabatically from 1050~MeV on to avoid going in the region where the approach is no longer valid. We take
\begin{equation}
	Gt(M_{\text{inv}})=Gt(M_{\text{cut}})e^{-\alpha(M_{\text{inv}}-M_{\text{cut}})}, \text{for}~M_{\text{inv}}>M_{\text{cut}},
\end{equation}
as done in Ref.~\cite{Debastiani:2016ayp}, where it was shown that moderate changes of $M_{\text{cut}}$ and $\alpha$ have only minor effects on the mass distributions.

\section{Results}\label{sec3}

	\begin{figure}
	\subfigure[]{
		\includegraphics[scale=0.5]{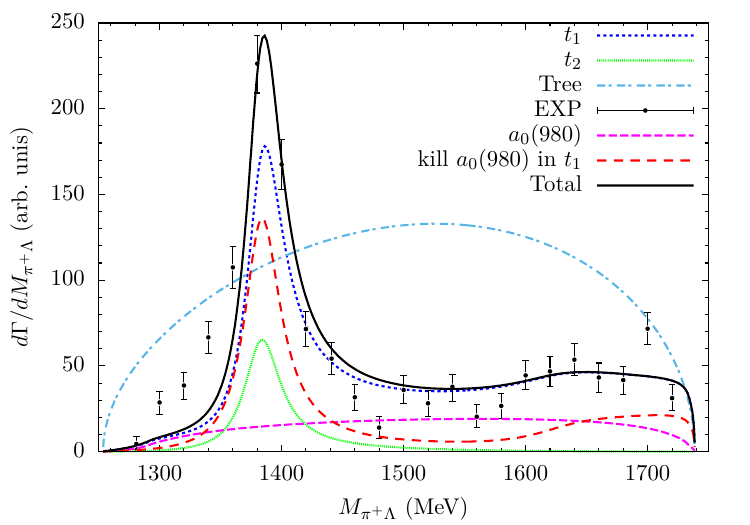}
	}
	\subfigure[]{
		\includegraphics[scale=0.5]{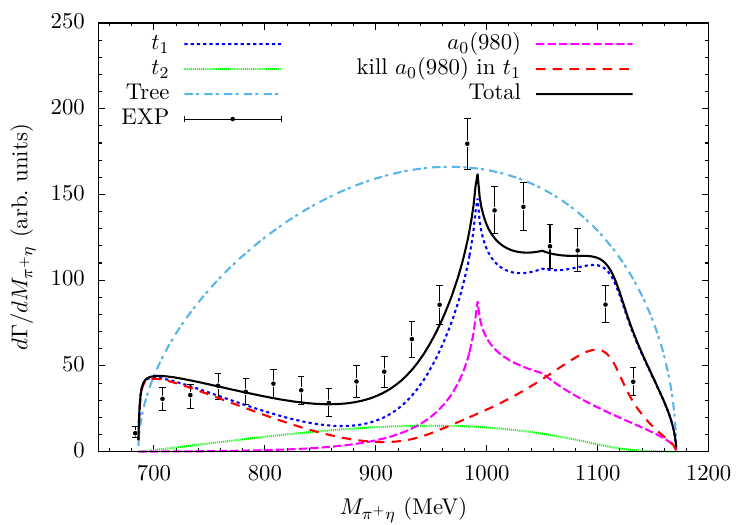}
	}
	\subfigure[]{
		\includegraphics[scale=0.5]{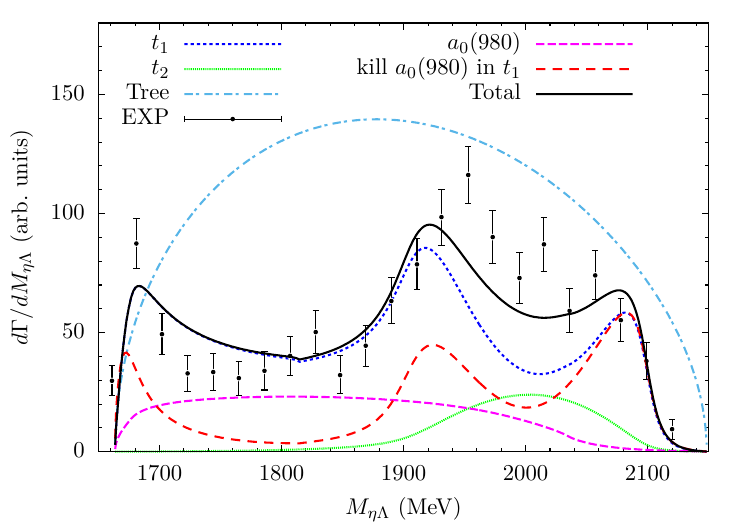}
	}
	\caption{The $\pi^+\Lambda$~(a), $\pi^+\eta$~(b), and $\eta\Lambda$~(c) invariant mass distributions of the $\Lambda^+_c\to \eta\pi^+\Lambda$ decay.}\label{fig:mass-distribuion-without-phase}
\end{figure}

\begin{figure}
	\subfigure[]{
		\includegraphics[scale=0.5]{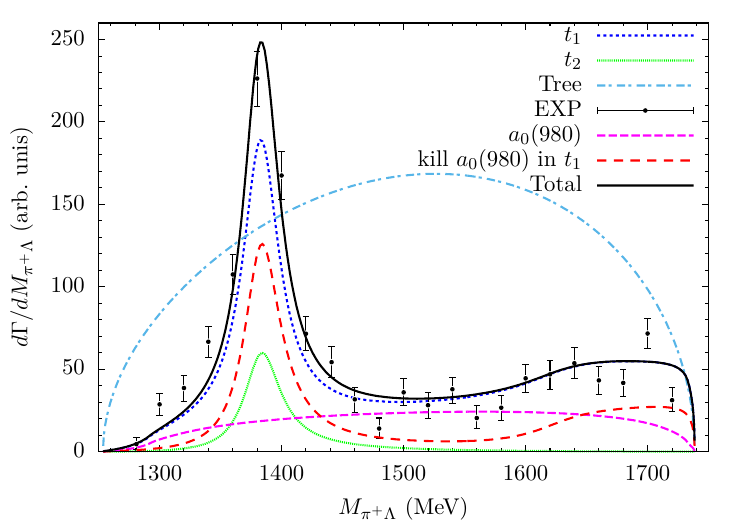}
	}
	\subfigure[]{
		\includegraphics[scale=0.5]{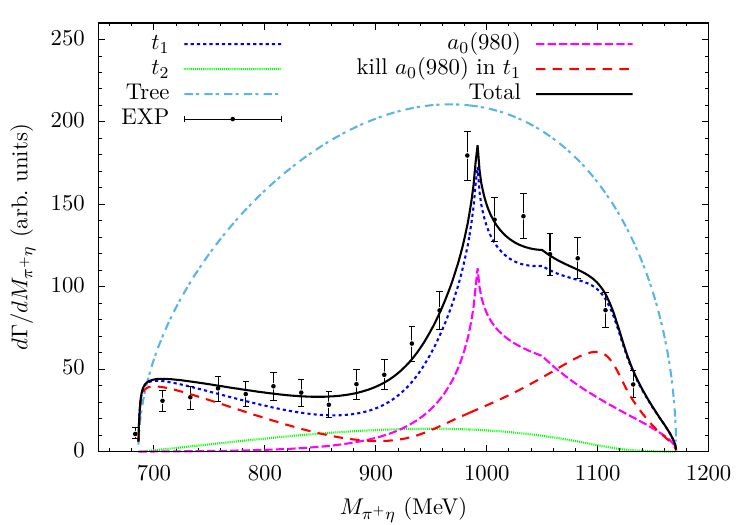}
	}
	\subfigure[]{
		\includegraphics[scale=0.5]{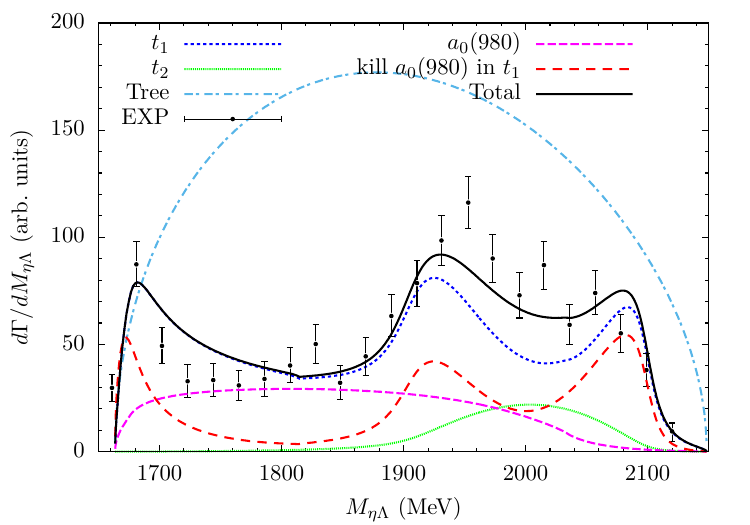}
	}
	\caption{The $\pi^+\Lambda$~(a), $\pi^+\eta$~(b), and $\eta\Lambda$~(c) invariant mass distributions of the $\Lambda^+_c\to \eta\pi^+\Lambda$ decay with $e^{i\phi}\beta$.}\label{fig:mass-distribuion-with-phase}
\end{figure}

In Fig.~\ref{fig:mass-distribuion-without-phase} we plot the mass distributions of $\pi^+\Lambda$, $\pi^+\eta$, and $\eta\Lambda$. We have taken into account that the experimental bins of masses are different for the three distributions in Ref.~\cite{BESIII:2024mbf}. The parameters of the fit are $A=0.457$, $\beta=0.098$, and $\chi^2/d.o.f.=208.47/ (63-2)=3.42$.

We observe at first glance that the agreement with the three experimental mass distributions is fair, and the basic features are reproduced. This is remarkable for an approach that relies upon only the free parameter $\beta$, up to the global normalization. In Fig.~\ref{fig:mass-distribuion-without-phase}(a) we see the $\pi^+\Lambda$ mass distribution, which is dominated by the $\Sigma(1385)$ excitation, and we see that both the spin non flip part (present in $t_1$) and the spin flip part present in $t_2$ contribute and produce the same shape in the $\pi^+\Lambda$ distribution. We also observe that the $a_0^+(980)$ contribution coming in our case from the $t_{\eta\pi^+,\eta\pi^+}$ amplitude, has a shape similar to that obtained in the analysis of Ref.~\cite{BESIII:2024mbf} but with smaller strength. The remarkable thing, however, is the weight of the tree level amplitude. We can see that by itself it would give a contribution to the mass distribution much bigger than what is measured by the experiment. This is telling us that there are strong interferences between the tree level, first term with $h_{\pi^+\eta\Lambda}$ in the $t_1$ amplitude of Eq.~(\ref{Eq:t1}), and the rest of the terms involving the $S$-wave interaction. The tree level is a very important term in the decay amplitude which is missed in the analysis of Ref.~\cite{BESIII:2024mbf}.

Next we look at Fig.~\ref{fig:mass-distribuion-without-phase}(b) with the $\pi^+\eta$ mass distribution. We see that we get a signal for $\Lambda a_0^+(980)$ production coming from the $t_{\eta\pi^+,\eta\pi^+}$ amplitude, which, however, has an integrated strength less than one half the one obtained in the analysis of Ref.~\cite{BESIII:2024mbf}. The $t_1$ amplitude removing the $a_0^+(980)$ contribution (the term with $t_{\eta\pi^+,\eta\pi^+}$) is due mostly to the $\Sigma(1385)$ contribution and has a similar shape to the one produced in the BESIII analysis (see Fig.~2 of Ref.~\cite{BESIII:2024mbf}). There is, however, a contribution from the spin flip $t_2$ amplitude which is absent in the BESIII analysis. Once again we see the large contribution of the tree level term by itself.

Finally, in Fig.~\ref{fig:mass-distribuion-without-phase}(c) we plot the result for the $\eta\Lambda$ mass distribution. Here we see that removing the $a_0^+(980)$ contribution to $t_1$ we get a contribution remarkable similar to the one produced by the $\Sigma(1385)$ in the BESIII analysis at $M_{\text{inv}}(\eta\Lambda)>1.8~\text{GeV}$, but now, included in $t_1$, coming from the terms with $t_{\eta\Lambda,\eta\Lambda}$, $t_{K^-p,\eta\Lambda}$, and $t_{\bar{K}^0n,\eta\Lambda}$, we observe that the peak of $\Lambda(1670)$ is automatically generated, while in the BESIII analysis it is a fitting term. The strength of the $\Lambda(1670)$ and $a_0^+(980)$ production are tied in our approach to the strength of the tree level and the $\pi^+K^-p$, $\pi^+\bar{K}^0n$ relative weight to the tree level provided by the weak production mechanism of Eq.~(\ref{Eq:H}). Once again, the tree level has a large strength here and very interesting, the spin flip part of the $t_2$ amplitude produces a bump around 2000~MeV, which is also absent in the BESIII analysis. One can trace back the different shape of the spin non flip and spin flip part of the $\Sigma(1385)$ production to the fact that the non spin flip part goes as $\text{cos}^2\theta$ (angle between $\vec{P}^*_{\pi^+}$ and $\vec{P}_{\eta}^*$) in the $|t_1|^2$ magnitude, see Eq.~(\ref{Eq:t1}), and $|t_2|^2$ goes as $\text{sin}^2\theta$, and we can see in Eq.~(\ref{Eq:ppidotpeta}) that $\text{cos}\theta$ is related to $M_{\text{inv}}(\pi^+\eta)$ and $M_{\text{inv}}(\pi^+\Lambda)$. Hence $\text{cos}^2\theta$ and $\text{sin}^2\theta$ have a very different shape when plotted in terms of the invariant masses.

While the fit obtained in terms of just one parameter, up to the global normalization, reproduces fairly well the features of all the mass distributions, we have taken advantage of the freedom in the phase of the $\beta$ term in Eq.~(\ref{Eq:t1}) and allow $\beta$ to acquire a phase going from 
\begin{equation}
	\beta\rightarrow e^{i\phi}\beta.
\end{equation}

In this case we have two free parameters and the fit returns the value $A=0.515$, $\beta=0.084$, $\phi=0.44\pi$ and $\chi^2/d.o.f.=125.81/ (63-3)=2.09$.

We plot the results in Fig.~\ref{fig:mass-distribuion-with-phase}. We see an improvement in the mass distributions and the $\chi^2$ value but the basic features were already obtained in the one parameter fit.

\section{Consideration of the internal emission mechanism}

\begin{figure}
	\centering
	\includegraphics[scale=0.65]{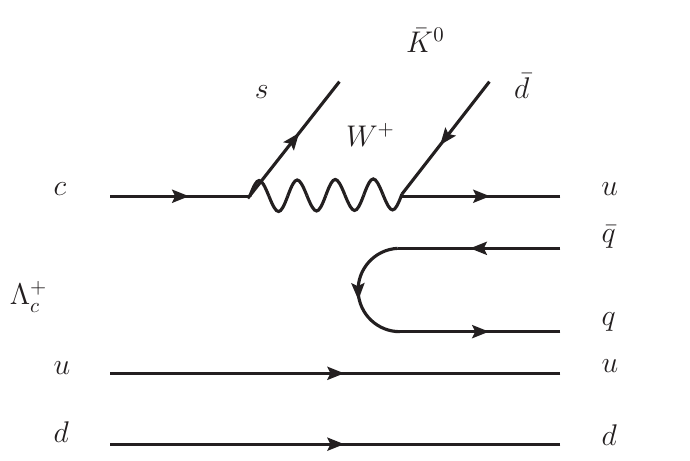}
	\caption{Mechanism for internal emission of quark level.}\label{3-body-internal}
\end{figure}

In addition, to the mechanisms based on external emission that we have considered, there is also a contribution from internal emission based on the mechanism depicted in Fig.~\ref{3-body-internal}.

After hadronization of the $uu$ and $ud$ quark pairs, we obtain
\begin{eqnarray}\label{Lambdac-internal-hadronization}
	&&\frac{1}{\sqrt{2}}c(ud-du)\chi_{MA}\rightarrow\frac{1}{\sqrt{2}}\bar{K}^0(u\bar{q}_iq_i(ud-du))\chi_{MA}\nonumber\\
	&=&\frac{1}{\sqrt{2}}\bar{K}^0\left\{\left(\frac{\pi^0}{\sqrt{2}}+\frac{\eta}{\sqrt{3}}\right)u(ud-du)+\pi^+d(ud-du)\right. \nonumber\\
	&&\left. +K^+s(ud-du)\right\}\chi_{MA} \nonumber\\
	&=&\frac{\bar{K}^0}{\sqrt{2}}\left\{\left(\frac{\pi^0}{\sqrt{2}}+\frac{\eta}{\sqrt{3}}\right)p+\pi^+n-\sqrt{\frac{2}{3}}K^+\Lambda\right\}\chi_{MA},\nonumber\\
\end{eqnarray}
where the results of Ref.~\cite{Miyahara:2016yyh} for $\phi_{MA}$ of the baryons have been used, together with the fact that the full wave function of the baryons is $1/\sqrt{2}(\phi_{MS}\chi_{MS}+\phi_{MA}\chi_{MA})$.\footnote{The factor $1/\sqrt{2}$ is missing explicitly in Refs.~\cite{Liu:2023jwo,Xie:2017erh}, but reabsorbed by a global normalization constant.}

\begin{figure}
	\subfigure[]{
		\includegraphics[scale=0.5]{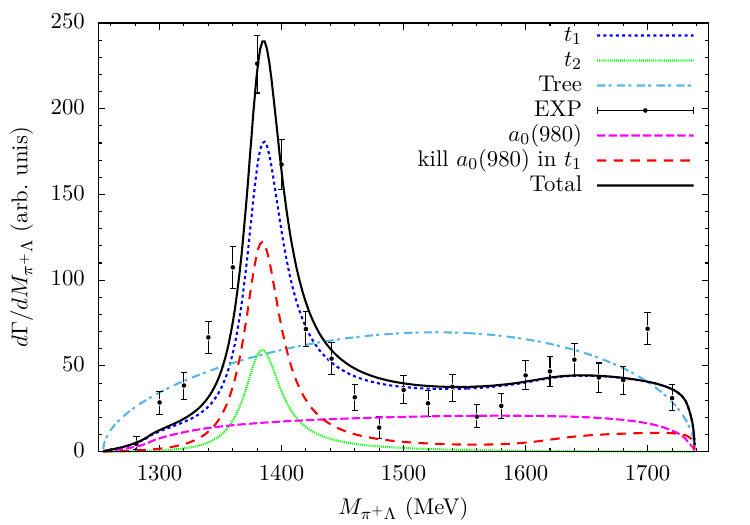}
	}
	\subfigure[]{
		\includegraphics[scale=0.5]{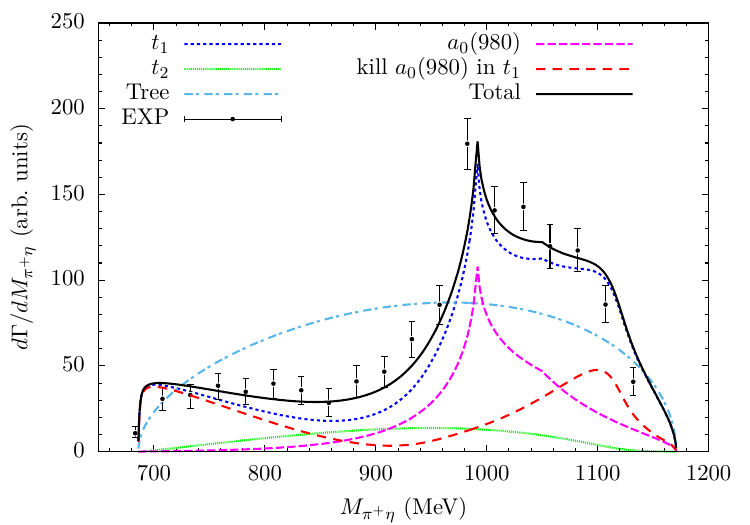}
	}
	\subfigure[]{
		\includegraphics[scale=0.5]{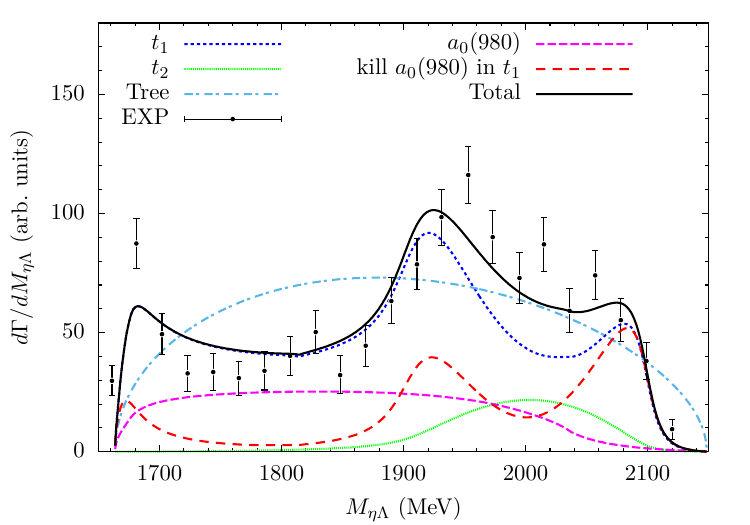}
	}
	\caption{The $\pi^+\Lambda$~(a), $\pi^+\eta$~(b), and $\eta\Lambda$~(c) invariant mass distributions of the $\Lambda^+_c\to \eta\pi^+\Lambda$ decay without the extra phase, including internal emission.}\label{fig:mass-distribuion-without-phase-add-gamma}
\end{figure}

\begin{figure}
	\subfigure[]{
		\includegraphics[scale=0.5]{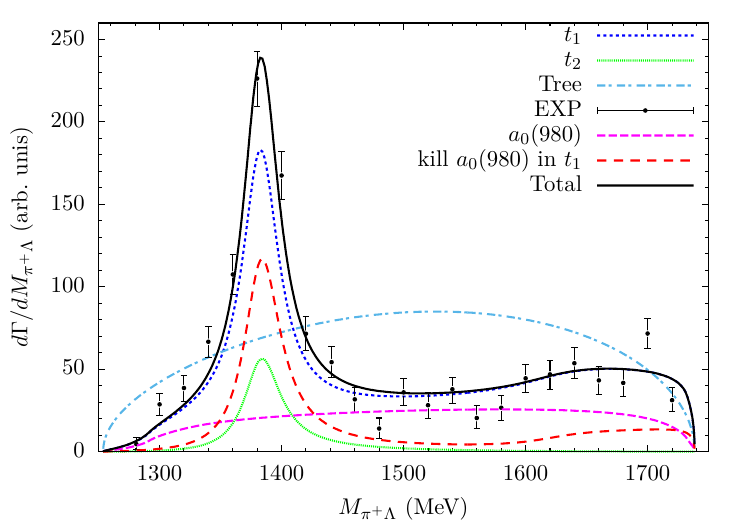}
	}
	\subfigure[]{
		\includegraphics[scale=0.5]{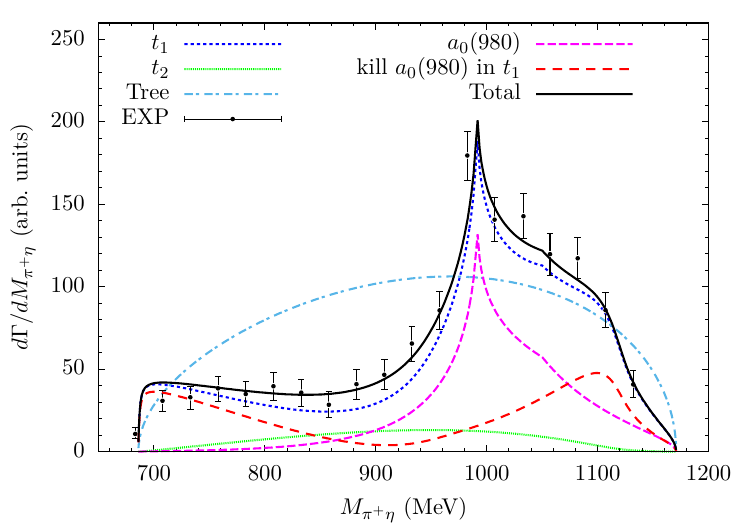}
	}
	\subfigure[]{
		\includegraphics[scale=0.5]{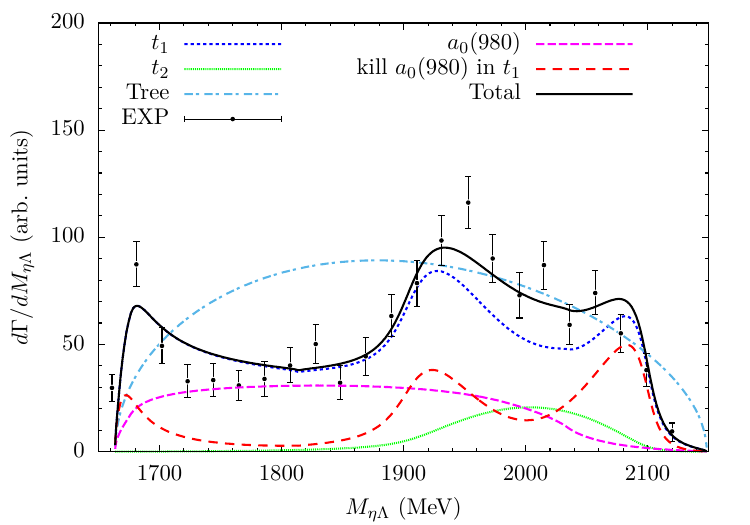}
	}
	\caption{The $\pi^+\Lambda$~(a), $\pi^+\eta$~(b), and $\eta\Lambda$~(c) invariant mass distributions of the $\Lambda^+_c\to \eta\pi^+\Lambda$ decay, including internal emission and with the extra phase for the $\Sigma(1385)$ term.}\label{fig:mass-distribuion-with-phase-add-gamma}
\end{figure}

We can see that now $\bar{K}^0K^+$ can interact and give $\pi^+\eta$. We easily account for this decay mode by adding to $t_1$ of Eq.~(\ref{Eq:t1}) the term
\begin{equation}\label{Eq:t-ie}
    t^{\text{ie}}=\gamma h_{\bar{K}^0K^+\Lambda}G_{\bar{K}^0K^+}(M_{\text{inv}}(\pi^+\eta))t_{\bar{K}^0K^+,\pi^+\eta}(M_{\text{inv}}(\pi^+\eta)),
\end{equation}
where
\begin{equation}
    h_{\bar{K}^0K^+\Lambda}=-\frac{1}{\sqrt{3}},
\end{equation}
and $\gamma$ is a relative weight of internal to external emission at the quark level, which should be about $1/N_c$ with $N_c$ the number of colors. We thus take $|\gamma|=1/3$ and see which sign is preferred. With this new input, we carry again a fit to the data and find that 
\begin{equation}
    \gamma=-\frac{1}{3},
\end{equation}
is clearly preferred. We show the new results of the fit in Figs.~\ref{fig:mass-distribuion-without-phase-add-gamma} and \ref{fig:mass-distribuion-with-phase-add-gamma}, with and without the extra phase to the $\Sigma(1385)$ term.

The results have not changed much by including the internal emission term of Eq.~(\ref{Eq:t-ie}). We can see that the strength of the $a_0$ at the peak has increased by about 20\% and the $\Lambda(1670)$ signal has been reduced 15\%, but the basic features are unaltered. The integrated strength of $a_0$ has been increased a bit and is now about 60\% of the one of the BESIII experiment~\cite{BESIII:2024mbf}, however, the shape is rather different than in the BESIII experiment.

We have conducted other fits related to the contribution of the internal emission. If we let $\gamma$ run free we also get a fit with $\gamma=-0.89$, but the fit cannot be considered better since the improvements in some parts of spectra come together with a reduction of the strength for the $\Lambda(1670)$ production. Also the fit depends much on the errors, since reducing artificially the errors in the $\Lambda(1670)$ region one produces a fit with $\gamma=-0.22$ but the valley from 1700$\sim$1800 MeV in $M_{\eta\Lambda}$ is worse reproduced.

We have made also fits fixing the angle $\phi$ to $90^{\circ}$ to reduce the interference with the $S$-wave terms and we get basically the same result as using $\gamma=-1/3$ and taking the angle free.

In summary, the different fits confirm what we obtain with the fits with $\gamma=-1/3$, which at the same time tell us that the basic features are already obtained with the external emission alone.

\section{ Conclusions }
We have studied the $\Lambda_c^+ \to \pi^+ \eta \Lambda$ decay observed by the Belle and BESIII Collaborations and conducted a theoretical study of the reaction to pin down the essential elements. Contrary to standard experimental analyses, where sums of resonance contributions with some incoherent background are fitted to the data, in our study we only introduce as a fit element the contribution of the $\Sigma(1385)$. In the experimental $\pi^+ \eta$, $\pi^+ \Lambda$ and $\eta \Lambda$ mass distributions one clearly sees structures tied to the  $a_0(980)$, $\Sigma (1385)$ and $\Lambda(1670)$ excitation. In our approach the $a_0(980)$ and $\Lambda(1670)$ resonances are dynamically generated from the $S$-wave meson meson and meson baryon interaction, respectively, and they come from rescattering mechanisms of the particles produced in a first step in the weak decay process, which are $\pi^+ K^- p$, $\pi^+ \bar{K}^0 n$ and $\pi^+ \eta \Lambda$. There is a tree level contribution, without rescattering, for the $\pi^+ \eta \Lambda$ channel, and the strength of this term is tied to the ones that generate the $a_0(980)$ and $\Lambda(1670)$ resonances. Once the fit to the data is done to determine the strength for $\Sigma(1385)$ excitation and the global normalization, the strength of the tree level and the $a_0(980)$ and $\Lambda(1670)$ resonances is automatically determined. This allows us to see that the strength of the tree level is very large. By itself it would produce an integrated width about two times bigger than the one observed experimentally. This means that there are large destructive interferences between the tree level and the $S$-wave rescattering terms. 
	
Another finding of the approach is the contribution of the spin flip part of the $\Sigma(1385)$ contribution, which is ignored in some theoretical papers and also in the BESIII experimental analysis. While the term appears with a strength of 1/4 with respect to the non spin flip part in the mass distributions, the different dependence on the invariant masses of these two terms, makes them to show up with different shapes in the $\pi^+ \eta$ and $\eta \Lambda$ mass distributions. 
	
The $a_0(980) \Lambda$ decay mode appears automatically in our approach due to the $\eta \pi^+$ rescattering. It is clearly visible in the results and in the $\eta \pi^+$  mass distributions, but it has an integrated strength about one half of that obtained from the BESIII analysis, solving a puzzle of an apparent contradiction with the visual effects in the Dalitz plot.   
	
We have also considered the contribution of the internal emission mode. The inclusion of this mode produced very small change in the mass distributions, a small increase in the $a_0(980)$ peak and a small decrease in the $\Lambda(1670)$ peak, without altering the basic feature of the reaction and the conclusions reached considering only the external emission mode.
	
In summary, we have shown that the consideration of the $a_0(980)$ and $\Lambda(1670)$ resonances as dynamically generated from the meson meson and meson baryon interactions, respectively, has allowed us to find a reasonable description of the invariant mass distributions for the $\Lambda_c^+ \to \pi^+ \eta \Lambda$ decay in terms of just one parameter, which is associated with the strength of the $\Sigma(1385)$ resonance, that appears in $P$-wave and is not linked to the other $S$-wave amplitudes. We also call the attention to the role played by the tree level contribution and the spin flip contribution in the $\Sigma(1385)$ excitation, which should in any case be looked up with attention in this and related reactions.

	
\section*{Acknowledgments}
This work is supported by the National Key Research and Development Program of China (No. 2024YFE0105200), the Natural Science Foundation of Henan under Grant No. 232300421140 and No. 222300420554, the National Natural Science Foundation of China under Grant No. 12475086, No. 12192263, No. 12175037, No. 12365019, No. 12335001, No. 12075288, No. 12435007, and No. 12361141819. 
This work is also supported by the Spanish Ministerio de Ciencia e Innovaci\'on (MICINN) under contracts PID2020-112777GB-I00, PID2023-147458NB-C21 and CEX2023-001292-S; by Generalitat Valenciana under contracts PROMETEO/2020/023 and CIPROM/2023/59. 
This work is partly supported by the Natural Science Foundation of Changsha under Grant No. kq2208257, the Natural Science Foundation of Hunan Province under Grant No. 2023JJ30647, the Natural Science Foundation of Guangxi Province under Grant No. 2023JJA110076; 
and also by the National Key R\&D Program of China under Grant No. 2023YFA1606703, the Youth Innovation Promotion Association CAS.

\end{document}